\documentclass[
 reprint,
 aps,
 superscriptaddress
]{revtex4-1}

\usepackage{graphicx}
\usepackage{amsmath}

\begin{document}

\title{
Emerging quasi-0D states at vanishing total entropy of the 1D hard sphere system:\\
a coarse-grained similarity to the car parking problem}
\author{Hiroshi Frusawa}
\email{frusawa.hiroshi@kochi-tech.ac.jp}
\affiliation{Institute for Nanotechnology, Kochi University of Technology, Tosa-Yamada, Kochi 782-8502, Japan.}

\begin{abstract}
A coarse-grained system of one-dimensional (1D) hard spheres (HSs) is created using the Delaunay tessellation, which enables one to define the quasi-0D state.
It is found from comparing the quasi-0D and 1D free energy densities that a frozen state due to the emergence of quasi-0D HSs is thermodynamically more favorable than fluidity with a large-scale heterogeneity above crossover volume fraction of $\phi_c=e/(1+e)=0.731\cdots$, at which the total entropy of the 1D state vanishes. 
The Delaunay-based lattice mapping further provides a similarity between the dense HS system above $\phi_c$ and the jamming limit in the car parking problem.
\end{abstract}

\maketitle

\section{Introduction}

Many decades have passed since van Hove's theorem \cite{vanHove,rigorous} was established regarding the non-existence of a phase transition in one-dimensional (1D) systems with short-ranged interactions.
Recently, the van Hove's theorem has been critically revisited with clarifying the three hypotheses behind this theorem \cite{rigorous}: homogeneity of particle distribution, no external fields, and the existence of hard-core potential.
As a consequence, the last two limitations have been removed in proving general non-existence theorem \cite{rigorous}, and yet the first restriction still needs to be imposed on the generalization.
In other words, it remains an open theoretical issue as to whether inhomogeneity found in disordered states, such as glassy phase, causes a highly-correlated state that is clearly distinguishable from homogeneous liquid assumed in these theorems.

In contrast, previous experimental and computer simulation results have reported several signatures of the local ordering and/or temporary freezing in dense 1D fluids \cite{giaquinta-gauss, ordering,theory-1,theory-2,experiments-1,experiments-2,peeters-1,peeters-2,bagchi,LJ,everlasting}.
Let us see below the details of the structural and dynamical indications in order to extract any clues to the understanding of relationship among heterogeneity, local ordering, and glassy dynamics.

The emergence of local ordering has been suggested by simulation studies.
On one hand, a latest simulation of the 1D Gaussian-core model, a typical soft-core system, has validated that particle clustering actually yields thermodynamic and structural anomalies \cite{giaquinta-gauss}.
Meanwhile, a set of Brownian simulations using various short-ranged interactions has demonstrated that the pronounced first peaks in the static structure factors are equally observed in dense 1D fluids where particles have a critical mobility \cite{ordering}.

Dynamically, the mean square displacement (MSD) of a tracer particle signals temporary freezing prior to crystallization at close packing.
The MSD has been found to obey the scaling law, $\sim t^{\alpha}$, with respect to a time interval $t$ \cite{theory-1,theory-2,experiments-1,experiments-2,peeters-1,peeters-2,bagchi,LJ,everlasting}.
Recent experimental and simulation studies have verified the crossover change of the MSD dependence from normal ($\alpha=1$) to single-file ($\alpha=0.5$) exponents \cite{experiments-1} with an increase of $t$.
Some experiments have also suggested the existence of a third region where the sub-diffusive exponent is further reduced to a long-time asymptote of $\alpha<0.5$ \cite{experiments-2}, and the diminished exponent of $\alpha<0.5$ has been reproduced by computer simulations of atomistic models \cite{peeters-1,peeters-2,bagchi}.
Furthermore, it was reported that initial conditions determine the long time limit of the tagged particle dynamics in an infinite 1D channel \cite{everlasting}.

\begin{figure}[hbtp]
        \begin{center}
	\includegraphics[
	width=7 cm
	]{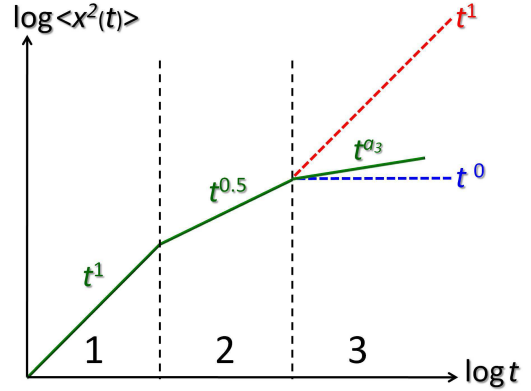}
	\end{center}
	\caption{Time dependences of the mean square displacement $\left<x(t)^2\right>\sim t^{\alpha}$ for a tagged particle. The green line is a schematic summary of the experimental and computer simulation results indicating three different exponents, $\alpha_1$, $\alpha_2$ and $\alpha_3$, in three time regimes of '1' to '3.' Our focus is on the third time regime in which two broken lines are delineated in addition to the green line. The red and blue lines correspond to the maximum and minimum exponents in the long time limit, respectively: $\alpha_{\infty}^{\mathrm{max}}\rightarrow1$ and $\alpha_{\infty}^{\mathrm{min}}\rightarrow0$.}
\end{figure}

Figure 1 outlines the above simulation and experimental results in a log-log plot of the MSD \cite{experiments-2,peeters-1,peeters-2,bagchi}.
The green line has three slopes, as delineated in Fig. 1, representing the experimental exponents of $\alpha_1=1$, $\alpha_{2}=0.5$, and  $\alpha_{3}< 0.5$.
In Fig. 1, we further compare the third exponent $\alpha_3$ with the final exponents in the long time limit: the maximum exponent, $\alpha_{\infty}^{\mathrm{max}}\rightarrow1$ (red line), and the minimum, $\alpha_{\infty}^{\mathrm{min}}\rightarrow0$ (blue line).
While the maximum exponent is associated with the Goldstone-like diffusion of the 1D chain as a whole \cite{peeters-2,borman-single}, the 1D particles confined in either a finite box or a harmonic potential will find their equilibrium positions, yielding the almost constant MSD (or the minimum exponent of $\alpha_{\infty}^{\mathrm{min}}\rightarrow 0$) \cite{box}.

It is thus seen from Fig. 1 that the sub-diffusive exponent approaches $\alpha_{\infty}^{\mathrm{min}}$ \cite{experiments-2,peeters-1,peeters-2,bagchi} due to the transient existence of a very large but finite box formed by transiently pinned particles, indicating that a tracer particle feels the temporary confinement in advance of recovering the normal exponent ($\alpha_{\infty}^{\mathrm{max}}\rightarrow 1$) associated with the Goldstone-like mode.
The suppression of global Goldstone-type flows should help constituent particles form a pseudo-crystalline state in a clustered chain, similarly to the confined particles that find their equilibrium positions in the end \cite{box}:
the structural indications, or local ordering, is ascribable to the dynamical signatures, or intermittent diffusion \cite{experiments-2,bagchi,LJ} in the 1D system, as well as in higher dimensional systems \cite{structure-dynamics}.
From the correspondence between statistics and dynamics in dense 1D fluids, we can infer that both of the structural and dynamical signatures found in experiments and simulations \cite{experiments-1,experiments-2,peeters-1,bagchi,LJ} arise from a generic phenomenon to crowded 1D systems unlike artifacts due to finite size effects and an artificial noise source \cite{peeters-2}, which actual systems are likely to involve.

Theoretically, the appearance of self-generated pinning by geometrically localized particles is to be explained in terms of collective particle dynamics.
The dynamical density functional theory \cite{ddft}, however, has found no signatures of the temporary freezing in the 1D HS system \cite{ddft,borman-original,borman-cluster,borman-review}, other than the dynamical crossover from the single-file diffusion to the Goldstone-type dynamics which can be reproduced with the help of the response function method using the frequency spectrum of relaxational dynamics in the 1D HS system \cite{borman-single,borman-review}.
From thermodynamic aspect, the exact free energy of the uniform 1D HS system has also predicted no anomaly without close packing as \cite{percus,tarazona, tonks,jepsen,parisi};
nevertheless, there exists a variety of theoretical criteria for a crossover phenomenon from the conventional liquid to a pseudo-crystalline state in the 1D HS system \cite{giaquinta-review}.

Consequently, we have a range of volume fraction $\phi_s$ as the onset densities of a solidlike behavior: $0.7< \phi_s< 0.83$ \cite{giaquinta-review}.
First, it is to be noted that a theoretical approach to the formation of high- and low-density 1D clusters has demonstrated the existence of a second maximum for the compressibility of an attractive 1D system at $\phi=0.71$, which is close to the aforementioned lower bound of $\phi_s$, and is also consistent with a large interatomic distance in monoatomic Au chains experimentally observed \cite{borman-cluster,borman-review}.
The upper bound of $\phi_s$ (the volume fraction of 0.83) is determined by the vanishing residual multi-particle entropy of the 1D HS system \cite{giaquinta-review,truskett,giaquinta-comment}, which also agrees with a jamming density of the square lattice model driven by strong external fields \cite{asep,pitard,new-shear}.
Yet the bases of these coincidences are tentative not merely because the decrease of attractive potential strength has been found to smear the extremal compressibility around the lower bound \cite{borman-cluster,borman-review}, but also beacause there remain some controversies over the aforementioned criterion for the upper bound using the residual entropy \cite{giaquinta-review,truskett,giaquinta-comment}.

In the 1D systems characterized by strong density fluctuations, temporarily generated clusters have a finite lifetime dependent on the packing density.
It follows that the theoretical treatments of repulsive 1D systems is likely to be hindered from detecting glass-like behaviors.
Accordingly, we need to take a fresh look at the inhomogeneous thermodynamics as a primary step toward the dynamical theory that is capable of exploring the emerging condition of intermittent dynamics caused by a global flow pinning.

\section{Our approach to the 1D thermodynamics}

We address the 1D inhomogeneity, or the local aperiodicity, using the fundamental measure theory (FMT), one of the density functional theories \cite{tarazona,fmt}.
The FMT has provided a successful description of the dimensional crossovers when a fluid is localized in a lower dimensionality, including the cases of narrow slits and cylindrical pores. The 3D free energy that is formulated by the FMT correctly converges to the exact 0D-limiting form, or the free energy of a cavity \cite{tarazona,fmt}.
According to the FMT, the fluid-solid transition in 3D HSs is interpreted as one related to dimensional crossover phenomena \cite{tarazona,fmt}.
Similarly, we adopt a dimensional crossover approach for determining a precursory volume fraction $\phi_c$ of crossover from the genuine 1D state to the coexistence of the 1D liquid state and the quasi-0D state (coarse-grained localization), assuming that the quasi-0D state specified below can be defined as a quasi-equilibrium state.

The remainder of this letter is organized as follows:
(i) First, the Delaunay tessellation \cite{delaunay1,delaunay2} of the original system is performed for creating a dual system of 1D HSs, which is further coarse-grained using the Delaunay cells.
Through the coarse-graining, the quasi-0D state is naturally introduced as an inactive state.
(ii) Next, we provide a thermodynamic criterion for the quasi-0D state.
(iii) Third, the quasi-0D free energy in the 1D HS system is formulated with the help of FMT.
(iv) As a result, we obtain the crossover volume fraction $\phi_c=e/(1+e)\approx 0.731\cdots$, at which the total entropy of the uniform (or coarse-grained) 1D HS system vanishes instead of the residual multi-particle entropy.
It is to be noted that the present density of $\phi_c$ does not only supports the lower bound in the above range of $\phi_s$, but is also close to the jamming density $\phi_j=0.747\cdots$ in the car parking problem \cite{parking1,parking2,parking3}.
(v) Last, we provide a realistic mapping that is based on the Delaunay tessellation of 1D HSs.
The Delaunay-based mapping relates the one-component HSs in continuous 1D space to a 1D lattice model of sites with continuously fluctuating densities, providing a coarse-grained similarity between the quasi-0D state in the HS system and the jammed state in the car parking problem.

\section{Setting up a dual system}

To extract the thermodynamics unique to the present freezing, we need to explore a minimal component of HSs that blocks the global flows of the 1D fluid.
We adopt the Delaunay tessellation of the 1D HS system as a method to create the unit cell because the Delaunay cells are determined by two-body correlations between nearest-neighbors, in contrast to the dual cells, the so-called Voronoi ones, which are not generated without knowing the three-particle configurations \cite{delaunay1}.
This section consists of two parts: we provide a dual system of the Delaunay cells, and define the inactive states as well as voids via coarse-graining.

\subsection{Delaunay cells}
Let $N$ be a given number of HSs in the 1D system.
Figure 2(a) shows that the $n$-th Delaunay cell on the right side is constituted by a pair of nearest-neighboring particles which are located, respectively, at $x_{n}$ and $x_{n+1}$ ($1\leq n\leq N$) and is specified by the center position of $s_n=(x_{n}+x_{n+1})/2$, the cell size of $w_n=|x_{n+1}-x_{n}|$, and its associated volume fraction:
   \begin{equation}
   \phi_n\equiv 1/w_n,
   \label{cell-density}
   \end{equation}
where we have set for brevity that $\sigma=1$ for the diameter $\sigma$ of the HSs.
The ensemble of Delaunay cell centers generates a 1D particle system, which is dual to the original 1D HS system.
When we consider a triplet of original HSs in contact with each other, the cell separation $d_n=|s_{n}-s_{n-1}|$ between the nearest-neighboring cells becomes equal to the particle diameter, $d_n=1$.
Correspondingly, the interaction potential $V(d_n)$ between nearest-neighboring cells takes the following form:
   \begin{equation}
   V(d_n)=
   \begin{cases}
   \infty&d_n<1\\
   0&d_n\geq 1,
   \end{cases}
   \label{interaction}
   \end{equation}
representing the same interaction as that of original HSs.

\subsection{Coarse-graining}

Our focus is on the emerging coexistence in space-time of active and inactive states.
The bimodal distribution of the dynamical activity was initially found in a larger variety of the kinetically constrained models of 1D to 3D lattices \cite{asep,pitard,new-shear,driven-garrahan,chandler-1,chandler-2,berthier,preprint,1d-coexistence}.
Subsequently, a 3D atomistic model of structural glass-formers obtained numerical evidence of the same type of dynamical coexistence \cite{structure-dynamics,chandler-science,chandler-prx,atom2,atom3}.
According to recent studies on the molecular dynamics simulations of the atomistic models \cite{chandler-science,chandler-prx,atom2,atom3}, excited dynamics of active particles is detected using a persistent particle displacement of length $a$.
At supercooled conditions, it has been found relevant to take the lower bound of $a$ to be of the order of a particle diameter in discriminating active dynamics from inactive states, or self-generated pinning.
Previous treatments of the dynamical coexistence thus require a coarse-graining of the particle systems.

\begin{figure}[hbtp]
       \begin{center}
	\includegraphics[
	width=6.5 cm
	]{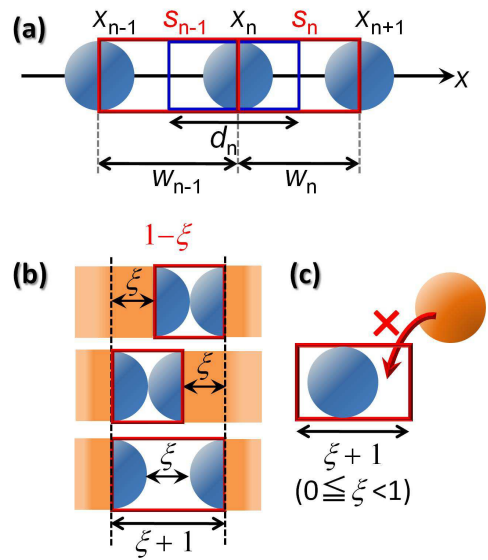}
	\end{center}
	\caption{(a) Three spheres are located at $x_{n-1}$, $x_n$, and $x_{n+1}$, creating two Delaunay cells (red frames) as well as one Voronoi cell (blue frame) with a width of $d_n$. While the center position of the left Delaunay cell with a width of $w_{n-1}$ is at $s_{n-1}$, that of the $w_n$-sized Delaunay cell on the right side is at $s_n$.
(b) A set of nearest-neighboring HSs creates a Delaunay cell whose mean center position is invariant under the vibrational motions as illustrated in schematic form. The top and middle configurations of the HS pairs represent synchronous fluctuations around the mean cell position. The bottom schematic, on the other hand, illustrates an asynchronous oscillation where the maximum gap length between the spheres is equal to $\xi$. (c) A blue sphere in a quasi-0D state trapped inside a cavity (red frame) that represents a maximum Delaunay cell. The range of $0\leq\xi<1$ excludes more than one particle.}
\end{figure}

In the Delaunay cell system generated by the 1D HSs, the inactive state is represented by an elementary process that one particle is blocked by the nearest-neighbor in a Delaunay cell, as a minimal unit of the vibrating confinement.
Various configurations of the particle pair forming the $n$-th Delaunay cell is shown in Fig. 2(b), from which it is found that the center position of the inactive cell fluctuates over the scale of maximum gap length $\xi$:
   \begin{eqnarray}
   &&\xi=a,\label{0d-gap}\\
   &&0\leq \xi<1,
   \label{gap}
   \end{eqnarray}
where we have adopted that $a=1$ following the previous treatments.

Equations (\ref{0d-gap}) and (\ref{gap}) indicate that where the inactive pinning state of a Delaunay cell can be mimicked by one-particle trapping in a quasi-0D cavity with a width of $\xi+1(\leq 2)$, as shown in Fig. 2(c) where we unite the halves of the constituent HS pair, while holding the cell size at its  maximum.
Equation (\ref{gap}) also reads that
   \begin{eqnarray}
   &&\phi_n\geq 1/(\xi+1)\equiv\phi_0,
   \label{density-corr}\\
   &&0.5<\phi_0\leq1
   \label{0d-phi}
   \end{eqnarray}
regarding the $n$-th cell density $\phi_n$ and the quasi-0D volume fraction, $\phi_0$, which is defined by the lower bound of $\phi_n$.
In terms of coarse-graining, the inactive state defined by eqs. (\ref{0d-gap}) to (\ref{0d-phi}) smears particle dynamics inside a quasi-0D box;
therefore, the inactive state is identified with the quasi-0D one.

Similarly, it is natural that active state having the condition $0\leq \xi<1$ provides a coarse-grained minimal volume fraction $\phi_1$ in the range of $0.5\leq\phi_1<1$.
For $\xi>1$, on the other hand, we set that $\phi_1=0$, because it is plausible to regard this type of Delaunay cell as a coarse-grained void, or a bubble, where a particle can be active without involving a cooperative motion \cite{chandler-2,void}.

\section{Thermodynamic criterion}

Before proceeding to the formulation of free energy, we consider a global condition that thermodynamically permits the coexistence between active and inactive particles as a metastable state \cite{langer} during the long-time relaxation from an inhomogeneous liquid with huge clusters to the uniform fluid having the minimum free energy.
In the following of this section, we first introduce the two-state model for specifying these states.
Then, we compare the total free energies similarly to the liquid-liquid phase separation phenomena, thereby establishing a global criterion.

\subsection{Two-state model}

We consider the two-state model that simplifies dense 1D systems using the densities as follows (see also Fig. 3):
while the active and inactive particles have the 1D and quasi-0D densities, $\phi_1$ and $\phi_0$, respectively, the inhomogeneous fluid consists of a collection of system-wide chains, or huge 1D clusters, with their smeared density of $\phi_1$, which necessarily creates empty gaps between the chains with those coarse-grained density $\phi_v$ set to be $\phi_v\approx0$.
The difference between the coexisting states is ascribed to the dilute regions which are constituted either by inactive quasi-0D particles having its local density of $\phi_0\geq 0.5$, or by voids of dynamically irrelevant particles with their actual volume fraction lower than 0.5.
Correspondingly, a small part of the quasi-0D particles confines the fluidic remainder, whereas the existence of voids allows clusters to experience the Goldstone-like diffusion.

The uniform density $\phi$ of Delaunay cells, which is smeared overall the system, is obtained from averaging these densities, similarly to phase separation systems.
Let $k$ be a number ratio of the inactive particles to the total, and $p$ denote a fraction of voids.
We have
    \begin{eqnarray}
    \phi&=&k\phi_0+(1-k)\phi_1\nonumber\\
    &=&p\phi_v+(1-p)\phi_1,
    \label{mean}
    \end{eqnarray}
with which the two-state model provides the average of the 1D and quasi-0D free energy densities, $f_1$ and $f_{0\xi}$, where $\xi$ represents the gap length between HSs;
$f_{00}$ signifies the free energy density due to a single Delaunay cell in the strictly 0D limit of $\xi\rightarrow0$.
We investigate below the stability of quasi-0D states in terms of the relationship between $f_1$ and $f_{0\xi}$, presuming that the quasi-0D state is thermodynamically definable \cite{langer}.

\subsection{A global view}
The free-energy density $f_{\mathrm{co}}$ of the coexisting state reads
   \begin{equation}
   f_{\mathrm{co}}(\phi;\phi_0,\phi_1)=(1-k)f_{0\xi}(\phi_0)+kf_1(\phi_1),
   \end{equation}
similar to the phase-separation free-energy.
For comparison, we consider the averaged free-energy density $f_{{\mathrm{het}}}$ of the above huge clusters, which is expressed as
   \begin{equation}
   f_{{\mathrm{het}}}(\phi;\phi_v,\phi_1)
   =(1-p)f_1(\phi_v)+pf_1(\phi_1),
   \end{equation}
supposing a large scale inhomegeneity based on the two-state model.

\begin{figure}[hbtp]
        \begin{center}
	\includegraphics[
	width=7.7 cm
	]{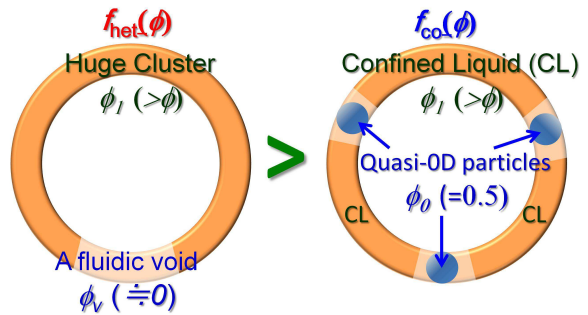}
	\end{center}
	\caption{A schematic representation of eq. (\ref{cross-necessary}). When a huge cluster with density $\phi_1$ larger than $\phi$ is formed and a tiny fraction of the remaining fluid is so dilute that the volume fraction $\phi_v$ is negligible, the free energy in the left system is approximately given by $f_{\mathrm{het}}(\phi;0,\phi_1)$. In the right system, on the other hand, quasi-0D particles create self-generated boxes, providing confined regions of the 1D fluid.}
\end{figure}

Comparison of these free energy densities provides a global criterion.
Figure 3 corresponds to the schematic representation, illustrating that the self-generated pinning is globally favored instead of maintaining the fluidity due to Goldstone-type flows that induces a large scale inhomogeneity.
Figure 3 reads that
   \begin{equation}
   f_{{\mathrm{het}}}(\phi;\phi_v=0,\phi_1)
   >f_{\mathrm{co}}(\phi;\phi_0,\phi_1),
   \label{cross-necessary}
  \end{equation}
which is fulfilled by the temporary freezing;
however, fluidity should be recovered in the end, because uniform 1D fluid has the minimum free energy: $f_{\mathrm{co}}(\phi)>f_1(\phi)$.

\section{Free energy}

The above discussions suggest that we could discriminate the self-generated pinning state in terms of the quasi-0D thermodynamics based on the Delaunay cell model.
Hence, we formulate the free energy per Delaunay cell in the quasi-0D state.

\subsection{Quasi-0D state}

Following the FMT, the quasi-0D free energy $F_{0\xi}(\eta)$ is given by the same form as the 1D free energy in the unit of thermal energy $k_BT$:
   \begin{equation}
   \beta F_{0\xi}(\eta)=\int dx\>f_{\mathrm{id}}+f_{\mathrm{ex}},
   \label{start}
   \end{equation}
where $\beta\equiv (k_BT)^{-1}$, the ideal part $f_{\mathrm{id}}$ per unit length, is given by $f_{\mathrm{id}}=\rho\ln \rho-\rho$, and the excess contribution $f_{\mathrm{ex}}$ is expressed as $f_{\mathrm{ex}}=-n_0\ln(1-n_1)$ \cite{tarazona,fmt}.
These forms use the localized particle density $\rho(x)$ and its weighted densities, $n_0$ and $n_1$, given as follows:
   \begin{equation}
   \rho_{0\xi}(x)=\frac{\eta}{\xi}
   \Theta\left(
   \frac{\xi}{2}-|x|
   \right),
   \label{cage_density}
   \end{equation}
$n_0(x)=(1/2)\int dx'\,\rho_{0\xi}(x')\,\delta\left(1/2-|x-x'|\right)$, and
$n_1(x)=\int dx'\,\rho_{0\xi}(x')\,\Theta\left(1/2-|x-x'|\right)$,  where we introduced the cavity-occupancy probability $\eta$ in the range of $0\leq\eta\leq 1$ \cite{tarazona,fmt}.
In the strict 0D-limit of $\xi\rightarrow 0$, the above densities, $\rho_{0\xi}$, $n_0$, and $n_1$, appropriately converge to the strict 0D concentrations, such as $\lim_{\xi\rightarrow 0}\rho_{0\xi}(x)=\eta\,\delta(x)$ in eq. (\ref{cage_density}), implying that the functional change from $\delta(x)$ to $(1/\xi)\Theta(\xi/2-|x|)$ expresses a coarse-grained cavity before close packing.

We then calculate the excess part of the free energy $f_{\mathrm{ex}}$ following the previous derivation in the FMT \cite{tarazona,fmt}.
Because the same integral relation,
$\int_{-\infty}^{\infty}dx\>n_0=-\int_1^{1-\eta}d(1-n_1)$,
as in the strict 0D-limit \cite{tarazona,fmt} applies to the quasi-0D form,
we are able to integrate $f_{\mathrm{ex}}$, yielding $\int dx\,f_{\mathrm{ex}}=(1-\eta)\ln(1-\eta)+\eta$, which is identical to the exact form of the excess 0D free-energy \cite{tarazona,fmt}.

While the ideal free-energy is straightforwardly given by
   \begin{equation}
   \int dx\,f_{\mathrm{id}}=\eta\ln(\eta/\xi)-\eta,
   \end{equation}
we calculate the excess part of the free energy $f_{\mathrm{int}}$ due to hard core interaction given by eq. (\ref{interaction}), following the previous derivation in the FMT \cite{tarazona,fmt}.
Because the same integral relation,
$\int_{-\infty}^{\infty}dx\>n_0=-\int_1^{1-\eta}d(1-n_1)$,
as in the strict 0D-limit \cite{tarazona,fmt} applies to the quasi-0D form,
we are able to integrate $f_{\mathrm{ex}}$, yielding
    \begin{equation}
    \int dx\,f_{\mathrm{int}}=(1-\eta)\ln(1-\eta)+\eta,
    \end{equation}
which is identical to the exact form of the excess 0D free-energy \cite{tarazona,fmt}.

Combining these results, eq. (\ref{start}) reads that
   \begin{eqnarray}
   \beta F_{0\xi}(\eta)
   =\eta\ln\eta+(1-\eta)\ln(1-\eta)-\eta\ln\xi.
   \label{reduce_free}
   \end{eqnarray}
Thus, we obtain the quasi-0D free energy per Delaunay cell having a single HS as
   \begin{eqnarray}
   \beta F_{0\xi}(\eta=1)&=&-\ln\xi\nonumber\\
   &=&\ln\left(\frac{\phi_0}{1-\phi_0}\right),
   \label{0d-cell-free}
   \end{eqnarray}
where use has been made of eq. (\ref{density-corr}) in the second line.
From eq. (\ref{0d-cell-free}), it follows that the quasi-0D free energy density $f_{0\xi}$ takes the following form
   \begin{eqnarray}
   f_{0\xi}(\phi_0)\equiv\frac{\beta F_{0\xi}(\eta=1)}{\xi+1}=
   \phi_0
   \ln\frac{\phi_0}{1-\phi_0},
   \label{0d_energy}
   \end{eqnarray}
where the range (\ref{0d-phi}) of $\phi_0$ is crucial in the following discussions.

\subsection{Comparison with the exact 0D state}

Let us compare eq. (\ref{reduce_free}) with the exact 0D free-energy \cite{fmt},
   \begin{equation}
   \beta F_{00}(\eta)=\eta\ln\eta+(1-\eta)\ln(1-\eta)+\eta\left\{\beta F_{00}(\eta=1)\right\},
   \label{original_0D}
   \end{equation}
where $\eta$ the average occupancy fraction.
Dividing the exact form $F_{00}(\eta)$ into the ideal and excess parts of $F_0^{\mathrm{id}}(\eta)$ and $F_0^{\mathrm{ex}}(\eta)$, we have $\beta F_0^{\mathrm{id}}(\eta)=\eta\ln\eta-\eta+\eta\left\{\beta F_{00}(\eta=1)\right\}$ and $\beta F_0^{\mathrm{ex}}(\eta)=(1-\eta)\ln(1-\eta)+\eta$, respectively.
These expressions imply the following:
the ideal part, $\beta F_0^{\mathrm{id}}(\eta=1)=\beta F_{00}(\eta=1)-1$, provides no information and the uncertainty of $\beta F_0^{\mathrm{id}}(\eta=1)$ remains in the exact form, though the excess free energy per particle is identical to the thermal energy, such that $\beta F_0^{\mathrm{ex}}(\eta=1)=1$.

In contrast, our free energy (\ref{reduce_free}) provides the one-particle free energy of $\beta F_{0\xi}(\eta=1)=-\ln\xi$.
The resulting form can be rederived by considering the background 1D system:
because the canonical partition function $Z_N$ of an $N$-particle 1D system is generally defined as $Z_N=(1/N!)\int dx\,e^{-\beta\mathcal{H}}$, we can immediately validate that $Z_1\equiv e^{-\beta F_{0\xi}(\eta=1)}=(1/1!)\int dx\,e^{0}=\xi$ due to no interaction energy ($\mathcal{H}=0$) in the one-particle system.


\begin{figure}[hbtp]
        \begin{center}
	\includegraphics[
	width=8.5 cm
	]{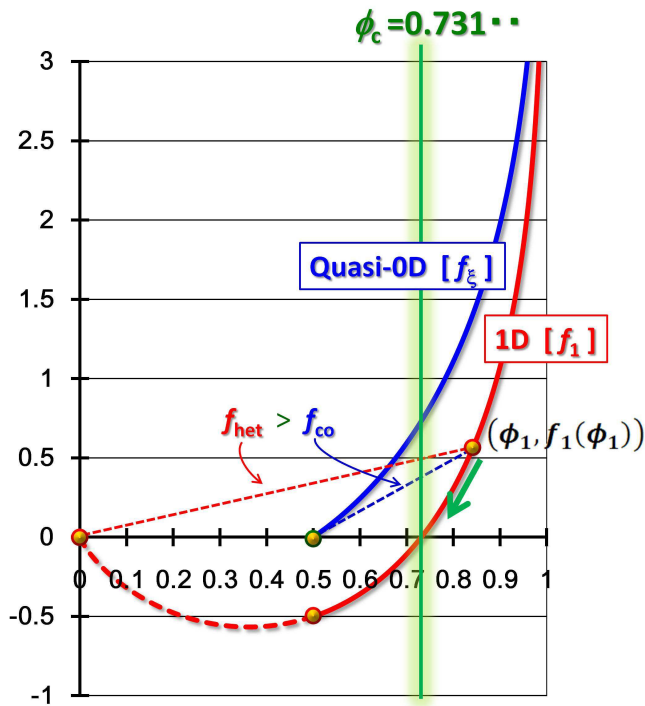}
	\end{center}
	\caption{The blue and red lines delineate the quasi-0D and 1D free-energies given by eqs. (\ref{0d_energy}) and (\ref{1d_energy}), respectively. The comparison between the blue and red broken lines demonstrates the equivalence of eqs. (\ref{cross-necessary}) and (\ref{slope}). The downward green arrow indicates that the slopes of $f_{\mathrm{co}}(\phi;0.5,\phi_1)$ and $f_{\mathrm{het}}(\phi;0,\phi_1)$ approach each other as $\phi_1\rightarrow\phi_c$ and merge at $\phi_1=\phi_c=0.731\cdots$.}
\end{figure}

\subsection{Comparison with the 1D fluid}

The exact free-energy of homogeneous 1D HSs is recovered by extending the use of $F_{0\xi}$ to the invalid range of $\xi\geq 1$.
In this case, the integration range with respect to $(1-n_1)$ must be changed from $\int_1^{1-\eta}d(1-n_1)$ to $\int_1^{1-\eta/\xi}d(1-n_1)$, due to the cage-size dependence of weighted density, such as $n_1(0)=\eta/\xi$.
Accordingly, we obtain the excess free energy, $\int dx\,f_{\mathrm{ex}}=(1-\eta/\xi)\ln(1-\eta/\xi)+\eta/\xi$, which is inconsistent with the 0D form of $(1-\eta)\ln(1-\eta)+\eta$ and vanishes in the limit of $\xi\rightarrow\infty$.

Thus, eq. (\ref{start}) leads to $\beta F_{0\xi\rightarrow\infty}(\eta=1)=-\ln\xi-1$, and we obtain the following forms of the 1D free-energy $F_1(\phi_1)$ per cell, the 1D free energy density $f_1(\phi_1)$ in the 1D HS fluid, from relating the gap length $\xi$ to the smeared density $\phi_1$ as $\phi_1=1/(\xi+1)$ \cite{percus,tarazona}:
  \begin{eqnarray}
   F_{0\xi\rightarrow\infty}(\eta=1)
   &=&F_1(\phi_1)=\ln\left(\frac{\phi_1}{1-\phi_1}\right)-1,\nonumber\\
   f_1(\phi_1)&=&-\frac{s(\phi_1)}{k_B}
   =\phi_1\left(
   \ln\frac{\phi_1}{1-\phi_1}-1\right),
   \label{1d_energy}
  \end{eqnarray}
where the free energy density $f_1$ in the unit of thermal energy $k_BT$ has been related to the total entropy density $s_1(\phi_1)$ as $f_1=-s_1/k_B$ because the free energy density $f_1$ of the HS system is determined only by the entropic contribution.

\section{Crossover density}

It is found from comparing the slopes of the two broken lines in Fig. 4 that eq. (\ref{cross-necessary}) reads the following inequality:
 \begin{equation}
  \frac{f_1(\phi_1)}{\phi_1}<
  \frac{f_1(\phi_1)}{\phi_1-0.5}.
  \label{slope}
  \end{equation}
Equation (\ref{slope}) is further reduced to
   \begin{equation}
   f_1(\phi_1)=\phi_1\ln\left(\frac{\phi_1}{1-\phi_1}\right)>0,
   \end{equation}
which provides the global condition for realizing Fig. 3, or eq. (\ref{cross-necessary}):
   \begin{equation}
   \phi_1>\frac{e}{1+e}\equiv\phi_c.
   \label{fai1}
   \end{equation}
Because the mean density $\phi$ is close to $\phi_1$ in the case of tiny fractions of both quasi-0D particles and voids ($k,p<<1$ in eq. (\ref{mean})), eq. (\ref{fai1}) implies that a crossover phenomenon occurs at
   \begin{equation}
   \phi=\phi_c=0.731\cdots,
   \end{equation}
which is to be compared with the jammed density $\phi_j=0.747\cdots$ in the car parking lot problem \cite{parking1,parking2,parking3}.
We predict that the quasi-0D particles, emergent as a nucleation in the 1D HS fluid above $\phi_c$, interrupt the Goldstone-like mode, or the cooperative diffusion of a system-wide string of 1D HSs, which causes intermittency (or a diminished exponent of $\alpha<0.5$ in terms of the MSD) followed by single-file diffusion.
It is to be noted that we have
   \begin{equation}
   f_1(\phi_c)=-\frac{s(\phi_c)}{k_B}=0,
   \label{entropy-phi}
   \end{equation}
verifying the above statement that the total entropy vanishes at $\phi_c$.
In terms of the cell free energy, on the other hand, eq. (\ref{entropy-phi}) reads that
   \begin{equation}
   F_{0\xi=1}(\eta=1)=F_1(\phi_c),
   \end{equation}
providing another explanation of $\phi_c$ as follows: the quasi-0D Delaunay cell is created because the free energy per cell is lowered by the self-generated pinning ($F_{\xi=1}<F_1(\phi)$ for $\phi>\phi_c$).

\section{Connection with the car parking lot problem: a lattice model}

Figure 5 demonstrates typical results of our lattice mapping from the particle configurations in Fig. 2(a).
Our mapping rule connects the lattice and atomistic models via coarse-graining, as follows.
Let $\mathcal{L}[s_n(t)]=i(t)$ be a function that assigns the number $i(t)$ of the present lattice site to the $n$-th Delaunay cell at an instantaneous position of $s_n(t)$.
At an initial time of $t_0$, we set that $\mathcal{L}[s_1(t_0)]=1$ and
   \begin{eqnarray}
   \mathcal{L}[s_{n}(t_0)]=\mathcal{L}[s_{n-1}(t_0)]+1
   \quad(1\leq d_{n}<2)
   \end{eqnarray}
using the Voronoi-cell separation of $d_{n}\equiv|s_{n}(t_0)-s_{n-1}(t_0)|$.
The same $n$-th particle moves to other position at an arbitrary time of $t$.
Then, we adopt the rule that
   \begin{eqnarray}
   \mathcal{L}[s_{n}(t)]=\mathcal{L}[s_n(t_0)]\quad
   (0\leq|s_{n}(t)-s_n(t_0)|<1),
   \end{eqnarray}
which is accomplished by imposing a constraint, given below, on the cell density.

\begin{figure}[hbtp]
        \begin{center}
	\includegraphics[
	width=6 cm
	]{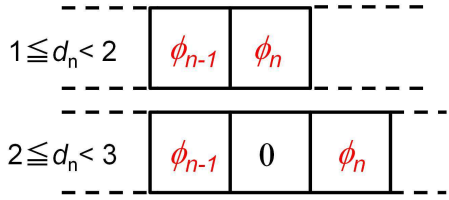}
	\end{center}
   \caption{Mapping from the configuration of three HSs in Fig. 2(a) to lattice systems in the case of $1\leq d_n<2$ and $2\leq d_n<3$.}
\end{figure}

The Delaunay densities, $\phi_{n-1}=1/w_{n-1}$ and $\phi_{n}=1/w_{n}$, of nearest-neighbor cells satisfy a reciprocal relation that
   \begin{equation}
   \phi_{n-1}^{-1}+\phi_{n}^{-1}=2d_n,
   \end{equation}
which reads the relation of $w_{n-1}+w_{n}\equiv2d_n$ between the widths of the Delaunay and Voronoi cells.
To exclude the existence of void cells (see the under panel in Fig. 5), we need to impose, on the separation $d_n$ between any pair of nearest-neighboring Voronoi cells, the following constraint:
   \begin{equation}
   1\leq \frac{1}{2}\left(
   \phi_{n-1}^{-1}+\phi_{n}^{-1}
   \right)< 2,
   \label{constraint}
   \end{equation}
thereby yielding a sequence of lattice sites as shown in the top panel of Fig. 5.
Equation (\ref{constraint}) states that the gap length between neighboring HSs is less than the HS diameter.
It follows that the site number of the Delaunay-based lattice model is equal to the number $N$ of constituent particles and is an invariant.

In the quasi-0D state defined by eqs. (\ref{0d-gap}) and (\ref{gap}), the kinetic constraint (\ref{constraint}) is not only naturally satisfied, but the lattice number is also identical to the cell species ($i\equiv n$), which is similar to the jammed state in the car parking model of random sequential adsorption \cite{delaunay2,parking1,parking2,parking3}.
The difference follows: vibrational motions in the HS system keep $\phi_n$ to be a variable even in the quasi-0D state, whereas each lattice density, in a jammed configuration of parked cars, takes a fixed value due to the quenched position;
therefore, the quasi-0D HSs in the 1D system provide realizable configurations of jammed states in the car parking problem, as clarified by the present lattice mapping.

\section{Concluding remarks}

We obtained the analytical form of quasi-0D free energy supposing that coarse-grained localization is thermodynamically discriminable from the 1D HS fluid.
The difference between the 1D and quasi-0D free energy forms enables one to provide a thermodynamic criterion for coexistence of 1D and quasi-0D states analogously to the phase separation condition, thereby predicting that a crossover phenomenon due to the emergence of quasi-0D state occurs at $\phi_c=e/(1+e)=0.731\cdots$ (or vanishing total entropy of the uniform 1D system).
It is to be noted that the present result of $\phi_c$ is within the previous density range ($0.7<\phi_s<0.83$) which was interpreted as an indicator of solidlike ordering \cite{giaquinta-review};
in this context, our treatment has revealed the underlying physics above $\phi_c$, or the lower bound of $\phi_s$, that the existence of a tiny fraction of self-pinned particles in the quasi-0D state is thermodynamically favored, rather than the long-standing fluidity that allows the Goldstone-like diffusions of huge clusters.

Since the quasi-0D state permits the vibrational motions within the amplitude of HS diameter, the temporary coexistence of active (1D) and inactive (quasi-0D) states, or the associated slow dynamics including the intermittency \cite{experiments-2,bagchi,LJ}, should become more obvious in a coarse-grained system, or a 1D lattice model.
As an example, we have introduced the Delaunay-based lattice model with a kinetic constraint of eq. (\ref{constraint}), so that the following similarity has been clarified:
long-lived quasi-0D HSs existing over a large scale represent the annealed state of quenched jamming in the 1D car parking model of the random sequential adsorption at the jamming density of $\phi_j=0.747\cdots(>\phi_c)$ \cite{parking1,parking2,parking3}.

In terms of the original atomistic model, it is indispensable for extracting a dynamical signature of the crossover phenomenon at $\phi_c$ to face the dynamical density functional theory \cite{ddft} based on the following equation:
\begin{eqnarray}
\frac{\partial\rho_1(x,t)}{\partial t}
=\frac{k_BT}{\gamma}\,\frac{\partial}{\partial x}\left[\,
\rho_1(x,t)
\frac{\partial}{\partial x}
\frac{\delta F_1\{\rho_1(x,t)\}}{\delta\rho_1(x,t)}
\right]
\label{DDFT}
\end{eqnarray}
where $\gamma$ is the friction coefficient, and the 1D free energy functional $F_1\{\rho_1(x,t)\}$ is expressed as $F_1\{\rho_1(x,t)\}=\int dxf_1[\rho_1(x,t)]$ using the free energy density $f_1$ given by eq. (\ref{1d_energy}).
In an approximation for deriving the phase-field-crystal model, eq. (\ref{DDFT}) is reduced to an interfacial equation regarding the density modulation $\psi(x,t)\equiv\rho_1(x,t)-\overline{\rho}_1$ from a 1D chain's mean density $\overline{\rho}_1$ \cite{pfc}:
\begin{eqnarray}
\frac{\partial\psi(x,t)}{\partial t}
&=&D(\overline{\rho}_1)\,
\frac{\partial^2\psi(x,t)}{\partial x^2}\nonumber\\
D(\overline{\rho}_1)
&=&D_f(1-\overline{\rho}_1\hat{c}),
\label{PFD}
\end{eqnarray}
where $D_f=k_BT/\gamma$ denotes the diffusion coefficient of free HSs, and the gradient expansion of $\hat{c}(\overline{\rho}_1)=C_0-C_2\nabla^2+\cdots$ corresponds to the Fourier transform of the direct corrrelation function in the ${\bf k}$-space: $c({\bf k};\overline{\rho}_1)=C_0+C_2{\bf k}^2+\cdots$.
While it has been demonstrated that eq. (\ref{PFD}) with a truncation of $\hat{c}\approx C_0$ is relevant to describing density fluctuations of a finite cluster formed in dense 1D HS system \cite{borman-original,borman-cluster,borman-review} and can be extended to two-component systems \cite{borman-two}, no dynamical crossover has been detected prior to close packing.
Our thermodynamic view, however, suggests the possibility that the interfacial fluctuations of 1D clusters would compete with the free energetic gain of the quasi-0D state when $\phi\geq\phi_c=0.731\cdots$, which has not been taken into account by eq. (\ref{PFD});
along this line, we could provide a more elaborate insight into the 1D dynamical crossover in the atomistic HS system by combining the dynamical theory and our thermodynamic result based on the two-state model \cite{future}.

\section*{Acknowledgement}

I am indebted to the anonymous referee's useful comments for the discussions from dynamical aspects.

\end{document}